\documentclass[12pt]{article}%
\usepackage{amssymb}
\usepackage{amsfonts}%
\usepackage{amsmath}%
\usepackage{graphicx}

\begin{document}

\author{Khireddine Nouicer(\thanks{E-mail: nouicer@fias.uni-frankfurt.de})\\Department of Physics and Laboratory of Theoretical Physics (LPTh),\\Faculty of Sciences, University of Jijel,\\Bp 98, Ouled Aissa, 18000 Jijel, Algeria }
\title{Coulomb potential in one dimension with minimal length:
a path integral approach}
\date{}

\maketitle

\begin{abstract}
We solve the path integral in momentum space for a particle in the field of the Coulomb potential in one dimension in
the framework of quantum mechanics with the minimal length  given by
 $( \Delta X)_{0}=\hbar
\sqrt{\beta}$, where $\beta$ is a small positive parameter. From the spectral
decomposition of the fixed energy transition amplitude we obtain the exact energy
eigenvalues and momentum space eigenfunctions.
\end{abstract}


\section{\bigskip Introduction}

The Planck  length defines a natural limit of the spacetime
resolution and its extremely small value $l_{p}\approx10^{-35} m$
renders the physical effects at this scale beyond the actual
accessible energies in particle accelerators. The main issue of
recently proposed models, like models with extra dimensions and
models with a minimal length scale, is to lower the Planck scale to
experimentally accessible energies. Recently, in a series of papers,
Kempf et al. \cite{kempf0,kempf1,kempf2,kempf3} and Hinrichsen and Kempf
\cite{kempf4} developed a quantum mechanics
based on a modified commutation relation $\left[  \hat{x}_{i},\hat{p}%
_{j}\right]  =i\hbar\left[  \left(  1+\beta\hat{p}^{2}\right) \delta
_{ij}+\beta^{\prime}\hat{p}_{i}\hat{p}_{j}\right]$, where $\beta$
and $\beta'$ are small parameters. This commutation relation leads
to a generalized uncertainty principle (GUP) which defines a non
zero minimal length in position. A minimal length in position can be
found in various contexts like string theory \cite{gross}, loop
quantum gravity \cite{garay}, and noncommutative field theories
\cite{douglas} and the Holographic Principle \cite{hooft,susskind}.
One major feature of the minimal length is that the physics below
this scale becomes inaccessible and then defining a natural cut-off
which prevents from the usual UV divergencies, and then put a limit
to the possible resolution of the spacetime. The other consequence
of the presence of the minimal length  is the appearance of an
intriguing UV/IR mixing. This mixing between UV and IR divergences
was also noticed in the ADS/CFT correspondence \cite{suskind,pet}
and in the canonical noncommutative quantum field theories
\cite{douglas}. On the other hand some scenarios have been proposed
where the minimal length is related to large extra dimensions
\cite{hos01}, to the running coupling constant \cite{hos02} and to
the physics of black holes production \cite{hos03}.

Recently, the effect of the minimal length in quantum mechanical
problems has been studied in several works. Among them, we cite: the
solution of the harmonic oscillator in arbitrary dimensions with
minimal length in the momentum space representation
\cite{kempf0,kempf1,minic}; the cosmological constant problem and
the classical limit of the physics with minimal length
\cite{chang,chang01}; the solution of the Coulomb potential in one
and three dimensions, has been studied respectively in
\cite{tkachuk} and \cite{brau,akhoury}; the Casimir force for the
electromagnetic field confined between parallel plates has been
computed in \cite{nouicer}, and the electron magnetism has been
investigated in \cite{nouicer1, nouicer2}. In this paper, we use
another approach, namely the path integral formalism in the momentum
space representation, to obtain the energy eigenvalues and the
momentum eigenfunctions of a particle in the field of Coulomb
potential in one dimension in the presence of the minimal length.
The physical interest in the 1D Coulomb potential comes from its
potential applications in semiconductors or insulators \cite{reyes}.

The rest of the paper is organized as follows. In the next section,
we introduce the main relations of quantum mechanics with minimal
length, such as maximally localization states, modified
orthogonality and completeness relations. In section 3, we extend
the path integral in the momentum space representation and the fixed
energy transition amplitude with regulating functions to time
independent systems in the presence of the minimal length . In
section 4 we determine, from the spectral decomposition of the fixed
energy transition amplitude, the exact energy eigenvalues and
momentum space eigenfunctions of a particle in the Coulomb potential
in one dimension. The last section is left for concluding remarks. \

\section{Quantum mechanics with minimal length}

Let us consider the following one dimensional momentum space
realization of the position and momentum operators \cite{kempf1} \
\begin{equation}
X=i\hbar(1+\beta p^{2})\frac{\partial}{\partial p},\qquad
P=p,\quad\beta\geqslant0.
\label{xp}
\end{equation}
These operators satisfy the following deformed commutator and GUP
\begin{equation}
\left[  X,P\right]    =i\hbar\left(  1+\beta p^{2}\right),\qquad
\left(  \Delta X\right)  \left(  \Delta P\right)    \geq\frac{\hbar}%
{2}\left[  1+\beta(\Delta P)^{2}\right]  .\label{UV}%
\end{equation}
A minimization of the saturate GUP with respect to $\left(  \Delta
P\right)  $ leads to a  minimal length given by
\begin{equation}
\left(  \Delta X\right)  _{0}=\hbar\sqrt{\beta}. \label{delta}
\end{equation}
At this stage we introduce a first quantized Hilbert space
equipped with the complete basis $\{|p\rangle\}$ given by
\begin{equation}
\int{Dp}|p\rangle\langle p|=\mathbf{1}. \label{closure}
\end{equation}
The hermiticity requirement of the position operator leads to the
following squeezed momentum measure
\begin{equation}
Dp=\frac{dp}{(1+\beta p^2)},
\end{equation}
and consequently we obtain
\begin{equation}
\langle p|p'\rangle=(1+\beta p^2)\delta(p-p').
\end{equation}
The appearance of the minimal length given by Eq.$(\ref{delta})$
leads to a loss of the notion of localized states in the position
space since we cannot probe the coordinates space with a
resolution less than the minimal length. However, the information
on position space is still accessible via the so called maximally
localization states \cite{kempf1}. These states are squeezed
states, which saturate the GUP and verify the constraint given by Eq.$\left(
\ref{delta}\right)$.

The maximally localization states denoted $\mid \psi _{\xi
}^{\text{ml}}>$ \ are defined as states localized around a
position $\xi $ \ such that we have
\begin{equation}
<\psi _{\xi }^{\text{ml}}\mid X\mid \psi _{\xi }^{\text{ml}}>=\xi ,\quad
\left( \Delta X\right) _{\mid \psi _{\xi }^{\text{ml}}>}=\left( \Delta
X\right) _{0},
\end{equation}%
and are solutions of the following equation
\begin{equation}
\left( X-\left\langle X\right\rangle +\frac{\left\langle \left[ X,P\right]
\right\rangle }{2\left( \Delta P\right) ^{2}}\left( P-\left\langle
P\right\rangle \right) \right) \mid \psi _{\xi }^{\text{ml}}>=0.
\end{equation}%
In the momentum space representation and with the expression of the position operator given by Eq.$\left( \ref{xp}%
\right)$ we obtain
\begin{equation}
\left( i\hbar \left( 1+\beta p^{2}\right) \frac{d}{dp}-\left\langle X\right\rangle
+i\hbar\frac{\left(1+\beta\langle P^2\rangle\right)}{2\left( \Delta P\right) ^{2}}%
\left( p-\left\langle P\right\rangle \right) \right) \psi _{\xi }^{\text{ml}%
}\left( p\right) =0.
\end{equation}%
The solution to this equation is given by
\begin{equation}
\psi _{\xi }^{\text{ml}}\left( p\right) =\frac{N}{\sqrt{2\pi \hbar }}\left(
1+\beta p^{2}\right) ^{-\frac{1+\beta \left( \Delta P\right) ^{2}+\beta
\left\langle P\right\rangle _{{}}^{2}}{4\beta \left( \Delta P\right) ^{2}}%
}e^{\left[ \frac{\left\langle X\right\rangle }{i\hbar \sqrt{\beta }}-\frac{%
\left( 1+\beta \left( \Delta P\right) ^{2}+\beta \left\langle P\right\rangle
^{2}\right) \left\langle P\right\rangle }{2\sqrt{\beta }\left( \Delta
P\right) ^{2}}\right] \tan ^{-1}\left( p\sqrt{\beta }\right) }.
\end{equation}%
We set $N=1$ by rescaling the states $N^{-1}\psi _{\xi
}^{\text{ml}}\left( p\right) \rightarrow \psi _{\xi
}^{\text{ml}}\left( p\right) .$ The states of absolutely maximal
localization are those with $\left\langle X\right\rangle =\xi ,$
$\left\langle P\right\rangle =0$ and, if we restrict these states
to the ones for which $\Delta P=1/\sqrt{\beta }$, we obtain
\begin{equation}
\psi _{\xi }^{\text{ml}}\left( p\right) =\frac{\left( 1+\beta p^{2}\right)
^{-\frac{1}{2}}}{\sqrt{2\pi \hbar }}\exp \left[ -\frac{i\xi }{\hbar \sqrt{%
\beta }}\tan ^{-1}\left( p\sqrt{\beta }\right) \right] .
\label{wave}
\end{equation}%
The states $\psi _{\xi }^{\text{ml}}\left( p\right) $ are physically
relevant ones with a finite energy
\begin{equation}
\left\langle \frac{p^{2}}{2m}\right\rangle _{\mid \psi _{\xi }^{\text{ml}}>}=%
\frac{1}{4\pi \hbar m}\int \frac{p^{2}dp}{\left( 1+\beta p^{2}\right) ^{3}}=%
\frac{1}{8\beta ^{\frac{3}{2}}\hbar m}.
\end{equation}%
On the other hand, the minimal length renders the maximally
localized states no longer orthogonal like the coherent states.
Indeed we show that
\begin{eqnarray}
<\psi _{\xi }^{\text{ml}}\mid \psi _{\xi ^{\prime }}^{\text{ml}}>&=&\frac{1}{%
2\pi \hbar }\int \frac{dp}{\left( 1+\beta p^{2}\right) ^{2}}\exp \left(
\frac{i\left( \xi -\xi ^{\prime }\right) }{\hbar \sqrt{\beta }}\tan
^{-1}\left( p\sqrt{\beta }\right) \right)  \notag \\
&=&\frac{2}{\pi \hbar \sqrt{\beta }}\frac{\sin \left( \frac{u\pi }{2}\right)}{u\left( u^2+4\right)}.
\end{eqnarray}%
where we have set $u=\frac{\left( \xi -\xi ^{\prime }\right) \pi }{\hbar
\sqrt{\beta }}.$

Finally, it is easy to verify that the set $\left\{ \mid \psi _{\xi }^{\text{ml%
}}>\right\} $ is complete in the sense that we have
\begin{equation}
\int d\xi \left( 1+\beta p^{2}\right) ^{2}\mid \psi _{\xi }^{\text{ml}%
}><\psi _{\xi }^{\text{ml}}\mid =\mathbf{1.}  \label{clos1}
\end{equation}

\section{Path integral in momentum space}

The path integral construction of the transition amplitude for quantum systems
with  minimal length in the momentum space representation
follows the well known canonical steps. In momentum space representation the
transition amplitude is given by%
\begin{align}
({p}_{b}T\mid{p}_{a}0_{a})  & =<{p}_{b}\mid U(T)\mid{p}_{a}>\nonumber\\
& =\lim_{N\longrightarrow\infty}<{p}_{b}\mid\prod\limits_{j=1}^{N+1}%
U(t_{j},t_{j-1})\mid{p}_{a}>,
\end{align}
with the infinitesimal evolution operator $U(t_{j},t_{j-1})=e^{-\frac{i}%
{\hbar}\epsilon\hat{H}\left(  t_{j}\right)  }$ and $\epsilon=\left(
t_{j}-t_{j-1}\right)  =\frac{T}{N+1}$. Inserting the completeness relation for
the momentum states given by Eq.(\ref{closure}) between each pair of
infinitesimal evolution operators gives%
\begin{equation}
(p_{b}T\mid p_{a}0)=\lim_{N\longrightarrow\infty}\prod\limits_{j=1}^{N}%
\int\frac{dp_{j}}{\left(  1+\beta p_{j}^{2}\right)  }%
\prod\limits_{j=1}^{N+1}(p_{j}t_{j}\mid p_{j-1}t_{j-1}),\label{propa}%
\end{equation}
where the infinitesimal transition amplitude is given by%
\begin{equation}
(p_{j}t_{j}\mid p_{j-1}t_{j-1})=<p_{j}\mid e^{-\frac{i}{\hbar}\epsilon\hat
{H}\left(  t_{j}\right)  }\mid p_{j-1}>.
\end{equation}
With the aid of the completeness relation for the maximally
localization states and Eq.(\ref{wave}), and assuming the standard
form of the Hamiltonian $\hat{H}=\frac
{\hat{p}^{2}}{2m}+\hat{V}(x)$, we arrive to the expression of the
infinitesimal transition
amplitude expressed as a phase space path integral%
\begin{align}
(p_{b}t_{b}\mid p_{a}t_{a})  & =\lim_{N\rightarrow
\infty}\prod\limits_{j=1}^{N}\int\frac{dp_{j}}{\left(  1+\beta p_{j}%
^{2}\right)  }\nonumber\\
& \times\prod\limits_{j=1}^{N+1}\int\frac{dx_{j}}{2\pi\hbar}\exp\left\{
\frac{i\epsilon}{\hbar}\left\{  \frac{x_{j}}{\epsilon\sqrt{\beta}}\left[
\tan^{-1}\sqrt{\beta}p_{j}-\tan^{-1}\sqrt{\beta}p_{j-1}\right]  -\frac
{p_{j}^{2}}{2m}-V\left(  x_{j}\right)  \right\}  \right\}  .\label{final}%
\end{align}
Now, following \cite{klein} we construct the path integral representation of the
fixed energy transition amplitude . The later is
defined by the following
matrix element%
\begin{equation}
(p_{b}\mid p_{a})_{E}^{f}=<p_{b}\mid\hat{R}(E)\mid p_{a}>
\end{equation}
of the resolvent operator%
\begin{equation}
\hat{R}\left(  E\right)  =\frac{i\hbar}{\hat{f}\left(  E-\hat{H}+i\eta\right)
}\hat{f},
\end{equation}
where $\hat{f}$ \ are regulating operators depending on $X$ and $P$. The
resolvent operator is defined by the Fourier transform of the time evolution operator%
\begin{equation}
\hat{R}\left(  E\right)  =\int_{0}^{T}dTe^{\frac{i}{\hbar}ET}\hat{U}\left(
T\right)  .
\end{equation}
Then, the path integral representation of the $f$-dependent fixed
energy
transition amplitude is given by%
\begin{equation}
(p_{b}\mid p_{a})_{E}^{f}=f\left(  0\right)  \int_{0}^{\infty}dT(p_{b}T\mid
p_{a}0)_{f},
\end{equation}
where $(p_{b}T\mid p_{a}0)_{f}$ is the transition amplitude associated with
the auxiliary Hamiltonian ${\hat{\mathcal{H}}}=\hat{f}(\hat{H}-E).$ The desired fixed
energy transition amplitude is given by
\begin{equation}
(p_{b}\mid p_{a})_{E}=\int_{0}^{\infty}dT(p_{b}T\mid p_{a}0)
\end{equation}
where $(p_{b}T\mid p_{a}0)$ is obtained by integrating out the
additional degree
of freedom $f$%
\begin{equation}
(p_{b}T\mid p_{a}0)=\int Df\Phi\left[  f\right]  (p_{b}T\mid p_{a}%
0)_{f},\label{f-indep}%
\end{equation}
along with the following condition%
\begin{equation}
\int Df\Phi\left[  f\right]  =1,
\end{equation}
with  $\Phi\left[  f\right]  $ a gauge fixing functional used to
select a specific gauge.

\section{Transition amplitude for the Coulomb potential in one dimension with
minimal length}

In this section we solve exactly the path integral of the Coulomb potential in one dimension
 in the presence of the minimal length. We start our calculation by substituting the potential $V(x)=-\alpha
/x$  in the path integral representation of the transition amplitude given by
Eq.(\ref{final}),
\begin{align}
(p_{b}T  & \mid p_{a}0)_{f}=\lim_{N\rightarrow
\infty}\prod\limits_{j=1}^{N}\int\frac{dp_{j}}{\left(  1+\beta p_{j}%
^{2}\right)  }\prod\limits_{j=1}^{N+1}\int\frac{dx_{j}}{2\pi\hbar}\nonumber\\
& \times\exp\left\{  \frac{i\epsilon}{\hbar}\left\{  \frac{x_{j}}%
{\epsilon\sqrt{\beta}}\left[  \tan^{-1}\sqrt{\beta}p_{j}-\tan^{-1}\sqrt{\beta
}p_{j-1}\right]  -f_{j}\left(  \frac{p_{j}^{2}}{2m}-E\right)  +f_{j}%
\frac{\alpha}{x_{j}}\right\}  \right\}  .
\end{align}
We shall work in the gauge given by
\begin{equation}
\Phi\left[  f\right]  =\prod\limits_{t}\frac{1}{x}\exp\left\{  -\frac
{i}{2\hbar x^{2}}\left[  f-x^{2}\left(  \frac{p^{2}}{2m}-E\right)  \right]
^{2}\right\} \label{gauge} .
\end{equation}
Inserting $\Phi\left[  f\right]$ in $(p_{b}T\mid p_{a}0)$ given by
Eq.(\ref{f-indep}) we obtain
\begin{align}
& (p_{b}T\mid p_{a}0)=\lim_{N\rightarrow\infty
}\prod\limits_{j=1}^{N}\int\frac{dp_{j}}{\left(  1+\beta p_{j}^{2}\right)
}\prod\limits_{j=1}^{N+1}\int\frac{dx_{j}}{2\pi\hbar}\prod\limits_{j=1}%
^{N+1}\int\frac{df_{j}}{\sqrt{2\pi\hbar}}\frac{1}{x_{j}}\nonumber\\
& \times\exp\left\{  \frac{i\epsilon}{\hbar}\left\{  \frac{x_{j}}%
{\epsilon\sqrt{\beta}}\left[  \tan^{-1}\sqrt{\beta}p_{j}-\tan^{-1}\sqrt{\beta
}p_{j-1}\right]  -\frac{1}{2}x_{j}^{2}\left(  \frac{p_{j}^{2}}{2m}-E\right)
^{2}-\frac{f_{j}^{2}}{2x_{j}^{2}}+f_{j}\frac{\alpha}{x_{j}}\right\}  \right\}
.
\end{align}
Performing the Gaussian integrals over $f$ $_{j}$ and $x_{j}$ we obtain%
\begin{align}
(p_{b}T  & \mid p_{a}0)={\left(  1+p_{b}^{2}%
/p_{E}^{2}\right)  ^{-1/2}\left(  1+p_{a}^{2}/p_{E}^{2}\right)  ^{-1/2}}%
\frac{\left(  2m/p_{E}^{2}\right)  }{\sqrt{2\pi i\epsilon\hbar}}%
\lim_{N\rightarrow\infty}\prod\limits_{j=1}^{N}\int\frac{(2m/p_{E}^{2})}%
{\sqrt{2\pi i\epsilon\hbar}}\frac{dp_{j}}{\left(  1+p_{j}^{2}/p_{E}%
^{2}\right)  \left(  1+\beta p_{j}^{2}\right)  }\nonumber\\
& \times\prod\limits_{j=1}^{N+1}\exp\left\{  \frac{i\varepsilon}{\hbar
}\left\{  \frac{2m^{2}}{\beta p_{E}^{4}\epsilon^{2}}\frac{\left[  \tan
^{-1}\sqrt{\beta}p_{j}-\tan^{-1}\sqrt{\beta}p_{j-1}\right]  ^{2}}{\left(
1+p_{j}^{2}/p_{E}^{2}\right)^2  }%
-\frac{\alpha^{2}}{2}\right\}  \right\}  .\label{pi02}
\end{align}

At this stage we expand the first term in the exponent around
$p_{j}$. This procedure leads to
\begin{equation}
\frac{2im^{2}}{\hbar\beta p_{E}^{4}\epsilon}\left[  \arctan p_{j}\sqrt{\beta
}-\arctan p_{j-1}\sqrt{\beta}\right]  ^{2}   =\frac{2im^{2}}{\hbar p_{E}%
^{4}\epsilon}\frac{(\Delta p_{j})^{2}}{(1+\beta p_{j}^{2})^2}\left\{
1-\frac{1}{6}\beta(\Delta p_{j})^{2}+\cdots\right\},
\end{equation}

where the omitted terms contain higher powers of $\Delta p_{j}$. Substituting in Eq.(\ref{pi02}), we obtain%
\begin{align}
(p_{b}T  & \mid p_{a}0)={\left(  1+p_{b}^{2}%
/p_{E}^{2}\right)  ^{-1/2}\left(  1+p_{a}^{2}/p_{E}^{2}\right)  ^{-1/2}}%
\frac{\left(  2m/p_{E}^{2}\right)  }{\sqrt{2\pi i\epsilon\hbar}}%
\lim_{N\rightarrow\infty}\prod\limits_{j=1}^{N}\int\frac{(2m/p_{E}^{2})}%
{\sqrt{2\pi i\epsilon\hbar}}\frac{dp_{j}}{\left(  1+p_{j}^{2}/p_{E}%
^{2}\right)  \left(  1+\beta p_{j}^{2}\right)  }\nonumber\\
&\times\prod\limits_{j=1}%
^{N+1}
\exp\Biggl\{\frac{i\varepsilon}{\hbar}\Biggl\{\frac{2m^{2}}{p_{E}%
^{4}\epsilon^{2}}\frac{(\Delta p_{j})^{2}}{\left(  1+p_{j}^{2}/p_{E}%
^{2}\right)^2(1+\beta p_{j}%
^{2})^2} -\frac{\frac{\beta m^{2}}{3p_{E}^{4}\epsilon^{2}}(\Delta
p_{j})^{4}}{\left( 1+p_{j}^{2}/p_{E}^{2}\right)^2(1+\beta p_{j}%
^{2})^2  }%
-\alpha^{2}+\cdots\Biggr\} \Biggr\}.\label{ampl}%
\end{align}
The correction terms $(\Delta p_{j})^{n}$ are calculated
perturbatively and replaced by their expectation values $<(\Delta p_{j})^{n}>$, using the following
formula \cite{klein}
\begin{equation}
<O(\Delta p)>=\int\frac{\sqrt{g(p)}}{\sqrt{2\pi
i\varepsilon\hbar/M}}e^{\frac{iM}{2\hbar\varepsilon}g_{mn}(p)(\Delta
p)^{m}(\Delta p)^{n}}O(\Delta p)d(\Delta
p).\label{average}%
\end{equation}
In our case, the metric $g(p)$ and the mass are given respectively
by
\begin{equation}
g_{mn}(p_j)=\frac{\delta_{mn}}{\left(  1+p_j^{2}/p_{E}^{2}\right)^2
(1+\beta p_j^{2})^2},\quad M=\frac{4m^{2}}{p_{E}^{4}}.
\end{equation}
Then, for the first non vanishing expectation values we easily obtain
\begin{align}
<(\Delta p_{j})^{2}&=&\left(  \frac{i\hbar\varepsilon}{4m^{2}/p_{E}^{4}%
}\right) g_{mn},\\
<(\Delta p_{j})^{4}>&=&3\left(  \frac{i\hbar\varepsilon}{4m^{2}/p_{E}^{4}%
}\right)  ^{2}g_{mn}^2,\\
<(\Delta p_{j})^{6}>&=&15\left(  \frac{i\hbar\varepsilon}{4m^{2}/p_{E}^{4}%
}\right)  ^{3}g_{mn}^3,
\end{align}
The remaining correction terms contains higher powers in
$\varepsilon$. Substituting in Eq.(\ref{ampl}), and considering only
the contributions which are relevant to order $\varepsilon$
we obtain%
\begin{align}
(p_{b}T  & \mid p_{a}0)={\left(  1+p_{b}^{2}%
/p_{E}^{2}\right)  ^{-1/2}\left(  1+p_{a}^{2}/p_{E}^{2}\right)  ^{-1/2}}\frac
{1}{\sqrt{2\pi i\epsilon\hbar/M}}\lim_{N\rightarrow\infty}\prod\limits_{j=1}%
^{N}\int\frac{1}{\sqrt{2\pi i\epsilon\hbar/M}}\frac{dp_{j}}{\left(
1+p_{j}^{2}/p_{E}^{2}\right)  \left(  1+\beta p_{j}^{2}\right)
}\nonumber\\
& \times\prod\limits_{j=1}^{N+1}\exp\left\{  \frac{i\varepsilon}{\hbar
}\left\{  \frac{M}{2\epsilon^{2}}\frac{(\Delta p_{j})^{2}}{\left(  1+p_{j}%
^{2}/p_{E}^{2}\right)^2 (1+\beta
{p}_{j}^{2})^2}+\beta\frac{\hbar^{2}p_{E}^{4}}{16m^{2}}-\alpha^{2}/2\right\}
\right\} \label{corr}%
\end{align}
In the following we adopt the mid-point prescription \cite{klein,kand}. To mate the
wild looking kinetic term we define a new path dependent time
$\sigma$ by the following symmetrized expression%
\begin{equation}
\varepsilon=\frac{\sigma_{j}}{(1+\beta p_{j}^{2})(1+\beta p_{j-1}^{2})}%
\end{equation}
and expand around the mid-point $\bar{p}_{j}=(p_j+p_{j-1})/2$ to obtain%
\begin{equation}
\varepsilon=\frac{\sigma_{j}}{(1+\beta\bar{p}_{j}^{2})^{2}}\left[  1-\frac
{1}{2}\beta\left(  \Delta p_{j}\right)  ^{2}+\mathcal{O}\left(  \beta
^{2}\right)  \right]  .
\end{equation}
Then, the exponent in Eq.$\left(  \ref{corr}\right)  $ reads as%
\begin{equation}
\hbox{Exp}=\frac{i}{\hbar}\left\{ \frac{M}{2\sigma_{j}}\frac{(\Delta
p_{j})^{2}}{\left( 1+p_{j}^{2}/p_{E}^{2}\right)^2 }\left[
1+\frac{1}{2}\beta\left(  \Delta p_{j}\right) ^{2}\right]
+\beta\sigma_{j}\frac{\hbar^{2}p_{E}^{4}}{16m^{2}}-\frac
{\alpha^{2}}{2}\sigma_{j}(1-2\beta\bar{p}_{j}^{2})+\mathcal{O}\left(
\beta^{2}\right)  \right\}  .
\end{equation}
There is another correction term proportional to $\left(  \Delta
p_j\right)  ^{2}$ coming from the measure. Using Eq.$(\ref{average})$ with the
following metric%
\begin{equation}
g_{mn}(p_j)=\frac{\delta_{mn}}{\left(
1+p_{j}^{2}/p_{E}^{2}\right)^2  },\label{metric2}%
\end{equation}
we replace $\left(  \Delta p_j\right)  ^{2}$ by%
\begin{equation}
<(\Delta p_{j})^{2}>=\left(  \frac{i\hbar\sigma}{4m^{2}/p_{E}^{4}%
}\right)  \left(  1+p_{j}^{2}/p_{E}^{2}\right)^2.
\end{equation}
Collecting all the three mentioned corrections terms we obtain%
\begin{align}
\hbox{Exp}  & =\frac{i\sigma_{j}}{\hbar}\left\{  \frac{M}{2\sigma_{j}^{2}%
}\frac{(\Delta p_{j})^{2}}{\left(  1+p_{j}^{2}/p_{E}^{2}\right)^2
 }+\left(  \frac{\beta\hbar^{2}p_{E}^{4}}{16m^{2}%
}\right)  _{1}-\left(  \frac{3\beta\hbar^{2}p_{E}^{4}%
}{16m^{2}}\right)  _{2}+\left(  \frac{\beta\hbar^{2}p_{E}^{4}}{8m^{2}}\right)
_{3}-\frac{\alpha^{2}}{2}(1-2\beta\bar{p}_{j}^{2})\right\} \nonumber\\
& =\frac{i\sigma_{j}}{\hbar}\left\{
\frac{M}{2\sigma_{j}^{2}}\frac{(\Delta p_{j})^{2}}{\left(
1+p_{j}^{2}/p_{E}^{2}\right)^2
}+\alpha^{2}\beta\bar{p}_{j}^{2}-\frac{1}{2}\alpha ^{2}\right\}  .
\end{align}
Here, we observe a cancelation of the corrections arising from the
time slicing. The same cancelation occurred in the path integral
treatment of the Coulomb potential in two and three dimensions
\cite{klein}. However, it is interesting to note that, even in one
dimension, the presence of the minimal length generates quantum
corrections similar to the quantum corrections generated by the
motion of point particles on curved spaces. This fact clearly
suggests some equivalence between the effects induced by space
curvature and the ones induced by the minimal length \cite{nieto}.

Now, let us incorporate the path dependent new time \ $T=\int_{0}^{\sigma_{b}}%
\frac{d\sigma}{\left(  1+\beta p^{2}\left(  \sigma\right)  \right)  ^{2}}$ by
means of the following identity%
\begin{equation}
\left[  \left(  1+\beta p^{2}{}_{b}\right)  (1+\beta p_{a}^{2})\right]
^{-1}\int_{0}^{\infty}d\sigma\delta\left(  T-\int_{0}^{\sigma_{b}}%
\frac{d\sigma}{\left(  1+\beta p^{2}\left(  \sigma\right)  \right)  ^{2}%
}\right)  =1.
\end{equation}
Then we have%
\begin{equation}
(p_{b}T\mid p_{a}0)=\frac{1}{2\pi\hbar}\int_{-\infty}^{+\infty}d\mathcal{E}%
e^{-\frac{i}{\hbar}\mathcal{E}T}(p_{b}\mid p_{a})_{\mathcal{E}},
\end{equation}
where $(p_{b}\mid p_{a})_{\mathcal{E}}$ is the fixed pseudo-energy to be
evaluated at $\mathcal{E}=0$ and is given by%
\begin{equation}
(p_{b}\mid p_{a})_{\mathcal{E}}={\left[  \left(  1+\beta p_{b}%
^{2}\right)  (1+\beta p_{a}^{2})\right]  ^{-1}}\int
_{0}^{\infty}d\sigma(p_{b}\sigma_{b}\mid p_{a}0),\label{fixed}%
\end{equation}
with%
\begin{align}
(p_{b}\sigma_{b}  & \mid p_{a}0)={\left(  1+p_{b}^{2}/p_{E}%
^{2}\right)  ^{-1/2}\left(  1+p_{a}^{2}/p_{E}^{2}\right)  ^{-1/2}}%
\frac{(2m/p_{E}^{2})}{\sqrt{2\pi i\sigma_{j}\hbar}}\lim_{N\rightarrow\infty
}\prod\limits_{j=1}^{N}\int\frac{(2m/p_{E}^{2})}{\sqrt{2\pi i\sigma_{j}\hbar}%
}\frac{dp_{j}}{\left(  1+p_{j}^{2}/p_{E}^{2}\right)  }\nonumber\\
& \times\prod\limits_{j=1}^{N+1}\exp\left\{  \frac{i\sigma_{j}}{\hbar}\left\{
\frac{2m^{2}}{p_{E}^{4}\sigma_{j}^{2}}\frac{(\Delta p_{j})^{2}}{\left(
1+p_{j}^{2}/p_{E}^{2}\right)^2   }%
+\alpha^{2}\beta\bar{p}_{j}^{2}-\frac{1}{2}\alpha^{2}+\mathcal{E}\right\}  \right\}.
\end{align}
To reduce this path integral to a more tractable one we use again
the mid-point prescription and calculate all the correction terms
proportional to $\left(  \Delta p\right)  ^{4}$ and $\left(
\Delta p\right)  ^{2}$ using Eq.$\left(  \ref{average}\right)  $
with $g$ \ given by Eq.$\left( \ref{metric2}\right)$. After a
lengthly and straightforward calculation we obtain
\begin{align}
(p_{b}\sigma_{b}  & \mid p_{a}0)={\left(  1+p_{b}^{2}/p_{E}%
^{2}\right)  ^{-1/2}\left(  1+p_{a}^{2}/p_{E}^{2}\right)
^{-1/2}}\frac{\left( 2m/p_{E}^{2}\right)  }{\sqrt{2\pi
i\epsilon\hbar}}\lim_{N\rightarrow\infty
}\prod\limits_{j=1}^{N}\int\frac{(2m/p_{E}^{2})}{\sqrt{2\pi i\sigma_{j}\hbar}%
}\frac{dp_{j}}{\left(  1+\bar{p}_{j}^{2}/p_{E}^{2}\right)  }\nonumber\\
& \times\prod\limits_{j=1}^{N+1}\exp\left\{
\frac{i\sigma_j}{\hbar}\left\{
\frac{2m^{2}}{p_{E}^{4}\sigma_{j}^2}\frac{(\Delta p_{j})^{2}}{\left(
1+\bar
{p}_{j}^{2}/p_{E}^{2}\right)  ^{2}}-\frac{\hbar^{2}p_{E}^{4}}{8m^{2}}%
+\alpha^{2}\beta\bar{p}_{j}^{2}-\frac{1}{2}\alpha^{2}+\mathcal{E}\right\}  \right\} ,
\end{align}
where $\frac{\hbar^{2}p_{E}^{4}}{8m^{2}}$ is the resulting final correction.

In the continuum limit we write Eq.(48) as%
\begin{equation}
(p_{b}\sigma_{b}\mid p_{a}0)=\frac{e^{-\frac{i}{2\hbar}(\alpha^{2}%
-2\mathcal{E)}\sigma}}{\left(  1+p_{b}^{2}/p_{E}^{2}\right)^{1/2}
\left(
1+p_{a}^{2}/p_{E}^{2}\right)^{1/2}}\int\mathcal{D}p\exp\left\{ \frac
{i}{\hbar}\int_{0}^{\sigma}d\sigma\left[
\frac{2m^{2}}{p_{E}^{4}}\frac {\dot{p}^{2}}{\left(
1+p^{2}/p_{E}^{2}\right)  ^{2}}+\alpha^{2}\beta
p^{2}-\frac{\hbar^{2}p_{E}^{4}}{8m^{2}}\right]  \right\}  .\label{correc01}%
\end{equation}
with $\mathcal{D}p=\frac{\left(  2m/p_{E}^{2}\right)  }{\sqrt{2\pi i\sigma
_{j}\hbar}}\lim_{N\rightarrow\infty}\prod\limits_{j=1}^{N}\int\frac{\left(
2m/p_{E}^{2}\right)  }{\sqrt{2\pi i\sigma_{j}\hbar}}\frac{dp_{j}}{\left(
1+\bar{p}_{j}^{2}/p_{E}^{2}\right)  }.$

Now, let us bring the kinetic term to the standard form by using
the following coordinate transformation
$p\in(-\infty,+\infty)\rightarrow\vartheta\in(-\pi p_{E},+\pi
p_{E})$
\begin{equation}
\vartheta_{j}=2p_{E}\arctan\frac{p}{p_{E}}.\label{trans}%
\end{equation}
Then, the transition amplitude is written as
\begin{equation}
(p_{b}\sigma_{b}\mid p_{a}0)=\frac{e^{-\frac{i}{2\hbar}(\alpha^{2}%
-2\mathcal{E)}\sigma}}{\left(  1+p_{b}^{2}/p_{E}^{2}\right)^{1/2}
\left(
1+p_{a}^{2}/p_{E}^{2}\right)^{1/2}}(\vartheta_{b}\sigma_{b}\mid\vartheta
_{a}0).\label{pp}%
\end{equation}
The transformation given by Eq.(\ref{trans}) generates a
correction term which cancels exactly the one in
Eq.(\ref{correc01}), and finally the transition amplitude can be
obtained as
\begin{equation}
(\vartheta_{b}\sigma_{b}\mid\vartheta_{a}0)=\int{\mathcal{D}}\theta
\exp\left\{  \frac{i}{\hbar}\int_{0}^{\sigma_{b}}d\sigma\left\{  \frac{M_{E}%
}{2}\dot{\theta^{2}}+\frac{\hbar^{2}}{8M_{E}p_{E}^{2}}\lambda\left(
\lambda-1\right)  \tan^{2}\frac{\theta}{2p_{E}}\right\}  \right\}
,\label{ppt}%
\end{equation}
with $\mathcal{D}\theta=\frac{1}{\sqrt{2\pi i\sigma_{j}\hbar/M_{E}}}%
\lim_{N\rightarrow\infty}\prod\limits_{j=1}^{N}\int\frac{d\theta_{j}}%
{\sqrt{2\pi i\sigma_{j}\hbar/M_{E}}}$,
$M_{E}=\frac{m^{2}}{p_{E}^{4}}$ and the parameter
$\lambda$ given by%
\begin{equation}
\lambda=\frac{1}{2}\left(  1+\sqrt{1+32\beta\frac{m^{2}\alpha^{2}}{\hbar^{2}}%
}\right) .
\end{equation}
The expression given by $\left(  \ref{ppt}\right)  $ is exactly the path
integral representation of the transition amplitude of a point particle in the symmetric Poschl-Teller
potential, whose spectral decomposition is given by \cite{kand}%
\begin{equation}
(\vartheta_{b}\sigma_{b}\mid\vartheta_{a}0)=\sum\limits_{n=0}^{\infty}%
A_{n}e^{-\frac{i}{\hbar}\frac{\hbar^{2}}{2M_{E}}\left[  n^{2}+(2n+1)\lambda
\right]  \sigma}\frac{1}{2p_{E}}\left(  \sin\frac{\theta_{b}}{2p_{E}}\sin
\frac{\theta_{a}}{2p_{E}}\right)  ^{\lambda}C_{n}^{\lambda}\left(  \cos
\frac{\theta_{b}}{2p_{E}}\right)  C_{n}^{\lambda}\left(  \cos\frac{\theta_{a}%
}{2p_{E}}\right),  \label{pt}%
\end{equation}
and where the normalization constant is
\begin{equation}
A_{n}=(\Gamma\left(  \lambda\right)  )^{2}\left[  \frac{2^{2\lambda-1}}{\pi
}\frac{n!\left(  n+\lambda\right)  }{\Gamma\left(  n+2\lambda\right)
}\right],
\end{equation}
and $C_n^{\lambda}(x)$ are the Gegenbauer polynomials.
Substituting Eqs.$(\ref{pt})$ and $(\ref{pp})$ in Eq. $\left(
\ref{fixed}\right) $, and integrating over $\sigma$ we obtain the
following spectral decomposition of
the fixed pseudo-energy transition amplitude%
\begin{equation}
(p_{b}\mid p_{a})_{\mathcal{E}}=\frac{i\hbar}{2p_{E}}\frac{\left[
\left( 1+\beta p_{b}^{2}\right)  \left(  1+\beta p_{a}^{2}\right)
\right] ^{-{1}}}{\left(  1+p_{b}^{2}/p_{E}^{2}\right)^{1/2}  \left(
1+p_{a}^{2}/p_{E}^{2}\right)^{1/2}}\sum\limits_{n=0}^{\infty}A_{n}%
\frac{\left(  \sin\theta_{b}\sin\theta_{a}\right)
^{\lambda}C_{n}^{\lambda }\left(  \cos\theta_{b}\right)
C_{n}^{\lambda}\left(  \cos\theta_{a}\right)
}{\frac{\hbar^{2}}{2M_{E}}\left[  n^{2}+(2n+1)\lambda\right]  -\frac
{\alpha^{2}p_{E}^{2}}{2}+2\mathcal{E}p_{E}^{2}}.
\end{equation}
Then, the fixed energy transition amplitude is simply obtained by
setting $\mathcal{E}=0$ in
the above expression%
\begin{equation}
(p_{b}\mid p_{a})_{E}=f\left(  0\right)
\frac{i\hbar}{2p_{E}}\frac{\left[ \left(  1+\beta p_{b}^{2}\right)
(1+\beta p_{a}^{2})\right]  ^{-1}}{ \left(
1+p_{b}^{2}/p_{E}^{2}\right)^{1/2}\left(
1+p_{a}^{2}/p_{E}^{2}\right)^{1/2}  }\sum\limits_{n=0}^{\infty}A_{n}%
\frac{\left(  \sin\theta_{b}\sin\theta_{a}\right)
^{\lambda}C_{n}^{\lambda }\left(  \cos\theta_{b}\right)
C_{n}^{\lambda}\left(  \cos\theta_{a}\right)
}{\frac{\hbar^{2}}{2M_{E}}\left[  n^{2}+(2n+1)\lambda\right]
-\frac {\alpha^{2}p_{E}^{2}}{2}}.
\end{equation}
Therefore, using the expression of $M_{E},$ we derive the following spectral condition%
\begin{equation}
\frac{\hbar^{2}p_{E}^{4}}{2m^{2}}[n^{2}+\left(  2n+1\right)  \lambda
]-\frac{\alpha^{2}p_{E}^{2}}{2}=0,\label{eq}%
\end{equation}
from which, using the expression of $\lambda$ and that
$p_{E}^{2}=-2mE_{n}$ \ for bound states, we obtain the expression
for the
energy spectrum%
\begin{equation}
E_{n}=-\frac{m\alpha^{2}}{2\hbar^{2}\left[  n^{2}+(n+\frac{1}{2})\left(
1+\sqrt{1+\left(  \frac{32m^{2}\alpha^{2}\beta}{\hbar^{2}}\right)  }\right)
\right]  },\quad n=0,1,2,...\label{energy}%
\end{equation}
It is obvious that for $\beta=0$ we recover the usual energy levels
of the Coulomb potential in one dimension \cite{physa}. The
corrections terms brought by the minimal length are obtained by
exploiting the fact that the $\beta$-dependent term in
Eq.$(\ref{energy})$ is a small quantity. Indeed, this term can be
written as $\left(\frac{(\Delta X)_0}{a_0}\right)^2$, where
$a_0=\frac{\hbar^2}{m\alpha}$ is the "Bohr radius" for the 1D Coulomb
atom. However, since the Bohr radius is the natural distance scale
for our system, and in order to be experimentally accessible it must
be greater than the minimal length. Then expanding Eq.(\ref{energy})
in terms of $\frac{(\Delta X)_0}{a_0}$ we obtain
\begin{equation}
E_{\tilde{n}}=-\frac{m\alpha^{2}}{2\hbar^{2}\tilde{n}^{2}}\left[
1-8\left(\frac{(\Delta X)_0}{a_0}\right)^2\frac{\left(  \tilde{n}+3/2\right)  }{%
\tilde{n}^{2}}\right]  ,\quad\tilde{n}=1,2,...\label{energyexpand}%
\end{equation}
We note that, besides numerical factors, we have reproduced the same
corrections terms as \cite{brau}. In Ref. \cite{akhoury} the correction term
proportional to $\left(  1/\tilde{n}^{3}\right)  $ is missed.

Then, let us turn to the spectral decomposition of the fixed
energy transition
amplitude. Indeed, we use the following relation%
\begin{equation}
\frac{i\hbar f\left(  0\right)  }{F(E)}\approx\frac{f\left(  0\right)
}{F^{\prime}(E_{n})}\frac{i\hbar}{E-E_{n}+i\eta},
\end{equation}
which when applied to $F(E)=2\hbar^{2}E^{2}[n^{2}+\left(  2n+1\right)
\lambda]+m\alpha^{2}E$ gives
\begin{equation}
\frac{i\hbar f\left(  0\right)  }{F(E)}\approx-\frac{f\left(  0\right)
}{m\alpha^{2}}\frac{i\hbar}{E-E_{n}+i\eta}.
\end{equation}
This result allows us to fix the unwanted arbitrary factor in Eq.$\left(
\ref{fixed}\right)  $ as $f\left(  0\right)  =m\alpha^{2}.$ With the aid of
the following relations%
\begin{align}
\cos\theta/2p_{E}  & =\frac{1}{\sqrt{1+p^{2}/p_{E}^{2}}},\\
\sin\theta/2p_{E}  & =\frac{p/p_{E}}{\sqrt{1+p^{2}/p_{E}^{2}}},
\end{align}
we finally obtain%
\begin{equation}
(p_{b}\mid p_{a})_{E}=\sum\limits_{n=0}^{\infty}\frac{i\hbar}{E-E_{n}+i\eta
}\Psi_{n}\left(  p_{b}\right)  \Psi_{n}\left(  p_{a}\right)  ,
\end{equation}
where  $E_{n}$ is the energy spectrum given by (\ref{energy}), and
$\Psi_{n}\left(  p\right)$ are the  normalized momentum space
eigenfunctions given  by
\begin{equation}
\Psi_{n}\left(  p\right)  =\frac{ e^{\frac{i\pi}{2}}\sqrt{\frac{A_{n}}{2p_{E}}} }{
\left(  1+\beta p^{2}\right)\left(  1+p^{2}/p_{E}^{2}\right)^{1/2}  %
}\left(  \frac{p/p_{E}}%
{\sqrt{1+p^{2}/p_{E}^{2}}}\right)  ^{\lambda}C_{n}^{\lambda}\left(  \frac
{1}{\sqrt{1+p^{2}/p_{E}^{2}}}\right)  .\label{onde}%
\end{equation}
Let us point, that although the deformed algebra in Eqs.(1-2) is only defined to first order in the minimal 
lenghth, the energy eigenvalues and momentum eigenfunctions are exact expressions. Let us also remark, that the energy eigenvalues are independent from the
parameter $\gamma$ which serves just to fix the measure in
the definition of the completeness relation$\left(
\ref{closure}\right)  .$ Therefore, it can be ignored in the
calculation or replaced by a more general term like $\gamma h(p)$
without affecting the physical quantities.

Finally, let us obtain the momentum space eigenfunctions of the 1D
Coulomb potential without the minimal length . Indeed, taking the
limit $\beta\rightarrow 0$ and
using the following relation \cite{grad}%
\begin{equation}
C_{n}^{1}\left(  \cos\theta\right)  =\frac{\sin\left(  n+1\right)  \theta
}{\sin\theta},
\end{equation}
we obtain, from $\left(  \ref{onde}\right),  $%
\begin{equation}
\Psi_{\tilde{n}}\left(  p\right)  =\frac{\sqrt{\frac{1}{4\pi p_{E}}}}{\left(
1+p^{2}/p_{E}^{2}\right)^{1/2} }\left[  e^{i\tilde{n}\arctan\frac{p}{p_{E}}%
}-e^{-i\tilde{n}\arctan\frac{p}{p_{E}}}\right]  .\label{nr}%
\end{equation}
This expression is different from the one obtained in \cite{physa},
where the authors of this paper reject the first term on the basis
that the coordinate space eigenfunction vanishes for $x\leq 0$, due
to the singular nature of the Coulomb potential for $x=0$. In our
case the singular point $x=0$ is now hidden by the minimal length,
since we cannot probe distances below the minimal length by virtue
of the GUP.

Finally, let us  point that the problems arising from the
noncommutativity of the position operators are absent in our one
dimensional framework, and that the generalization to higher
dimensions is not a simple task. In the case of higher even
dimensional spacetimes in the canonical noncommutativity, the path
integral construction of the generating functional has been done
in close analogy with the commutative case \cite{mangano}.

\section{Conclusion}

In summary, we have shown that the path integral in the momentum space representation of 1D Coulomb potential in the presence of a minimal length remains exactly
solvable. Using the mid-point expansion and the space-time
transformations, we mapped the problem to the one of a point
particle in the symmetric Poschl-Teller potential. Although the deformed algebra given by Eq.(1-2) is only defined to first order in the minimal length $(\Delta X)_0$, we have obtained from the
spectral decomposition of the fixed energy amplitude exact expressions of the energy eigenvalues and
momentum space eigenfunctions. A particular feature of our
calculation is the generation of quantum corrections, although we
considered a one dimensional model. These quantum corrections
arise naturally from the particle motion on curved spaces. This
property of the minimal length reveals the rich structure of the
spacetime, even in one dimension. Finally, we have checked the
correctness of our results by rederiving the results of the usual
Coulomb potential in one dimension and by the same way, we have
proved that the energy levels are proportional to $1/n^{2}$,
exactly like Coulomb potential in two and three dimensions. The
more important case of the Coulomb potential in three dimensions
with minimal length is under investigation and will be published
elsewhere.


\begin{thebibliography}{99}                                                                                               %
\bibitem {kempf0}A. Kempf, J. Math. Phys. {\bf 35}, 4483 (1994) \

\bibitem {kempf1}A. Kempf, G. Mangano and R. B. Mann, Phys. Rev. D
{\bf 52} 1108 (1995)

\bibitem {kempf2}A. Kempf, J. Phys. A {\bf 30}, 2093 (1997)

\bibitem{kempf3} A. Kempf and G. Mangano, Phys. Rev. D {\bf 55}, 7909 (1997)

\bibitem {kempf4}H. Hinrichsen and A. Kempf, J. Math. Phys. {\bf 37}, 2121 (1996)

\bibitem{gross} D. J. Gross and P. F. Mende, Nucl. Phys. B {\bf 303}, 407 (1998); D. Amati, M. Ciafaloni, and G. Veneziano, Phys. Lett. B {\bf 213}, 41 (1989); R. Lafrance and R. C. Myers, Phys. Rev. D {\bf 51}, 2584 (1995)

\bibitem{garay} L. J. Garay, Int. J. Mod. Phys. A {\bf 10}, 145 (1995)

\bibitem{douglas} M. R. Douglas and N. A. Nekrasov, Rev. Mod. Phys. {\bf 73}, 977 (2001); S. Minwalla, M. Van Raamsdonk, and N. Seiberg, J. High Energy Phys. 0002, 020 (2000); R. J. Szabo, Phys. Rep. {\bf 378}, 207 (2003)

\bibitem {hooft}G. 't Hooft, [gr-qc/3910026]

\bibitem {susskind}L. Susskind, J. Math. Phys.  {\bf 36}, 6377 (1995)

\bibitem {suskind} L. Susskind and E. Witten, [hep-th/9805114]

\bibitem {pet}A. W. Pet and J. Polchinski, Phys. Rev. \textbf{D 59, }065011 (1999)

\bibitem {hos01}S. Hossenfelder, Mod. Phys. Lett. A \textbf{19}, 2727 (2004)

\bibitem {hos02}S. Hossenfelder, Phys. Rev. D \textbf{70}, 1054003 (2004)

\bibitem {hos03}S. hossenfelder, Phys. Lett. B \textbf{598}, 92 (2004)

\bibitem {minic}L. N. Chang, D. Minic, N. Okamura and T. Takeuchi, Phys. Rev.
\textbf{D 65}, 125027 (2002)

\bibitem {chang}L. N. Chang, D. Minic, N. Okamura and T. Takeuchi, Phys. Rev.
\textbf{D 65}, 125028 (2002)

\bibitem {chang01}S. Benczik, L.N. Chang, D. Minic, N. Okamura, S. Rayyan and
T. Takeuchi, Phys. Rev. D {\bf 66}, 026003 (2002)

\bibitem{tkachuk} T. V. Fityo, I. O. Vakarchuk and V. M. Tkachuk, J. Phys. A: Math. Gen. {\bf 39}, 2143 (2006)

\bibitem {brau}F. Brau, J. Phys. A {\bf 32}, 7691 (1999)

\bibitem {akhoury}R. Akhoury and Y. -P. Yao, Phys. Lett. B {\bf 572}, 37 (2003)

\bibitem {nouicer}Kh. Nouicer, J. Phys. A {\bf 38}, 10027 (2005)

\bibitem{nouicer1} Kh. Nouicer, J. Math. Phys. {\bf 47}, 122102 (2006)

\bibitem{nouicer2} Kh. Nouicer, J. Phys. A: Math. Theor. {\bf 40}, 2125 (2007)

\bibitem {reyes}J. A. Reyes and M. del Castillo-Mussot, J. Phys. \textbf{A
32}, 2017 (1999)

\bibitem {klein}H. Kleinert, \textit{Path Integrals in quantum Mechanics,
Statistics, and Polymer Physics}, World Scientific (1990)

\bibitem {kand}D. C. Khandekar, S. V. Lawande and K. V. Bhagwat, \textit{Path
Integrals Methodes and their Applications}, World Scientific (1993)

\bibitem{nieto} L.M. Nieto, et al., Mod. Phys. Lett. A {\bf 14}, 2463 (1999)


\bibitem {grad}I. S. Gradshteyn and I. M. Ryzhik, \textit{Tables of Integrals,
Series and Produc}ts, Academic Press (1980)

\bibitem {physa}Y. Ran, L. Xue, S. Hu and R-K. Su, J. Phys. A {\bf 33},
9265 (2000)

\bibitem {mangano}G. Mangano, J. Math. Phys. {\bf 39}, 2584 (1998)

\end{thebibliography}
\end{document}